\documentclass[aps,showpacs,superscriptaddress,twocolumn]{revtex4}
\usepackage{makecell}

\usepackage[intlimits]{amsmath}
\usepackage{amssymb}
\usepackage{graphicx}
\usepackage{booktabs}
\usepackage{mathrsfs,slashed}
\usepackage{hyperref}
\usepackage{epsfig}
\usepackage{color}

\newcommand{\bea}{\begin{eqnarray}}
\newcommand{\eea}{\end{eqnarray}}

\newcommand{\nn}{\nonumber}

\def\lsim{\stackrel{\scriptstyle <}{\phantom{}_{\sim}}}
\def\gsim{\stackrel{\scriptstyle >}{\phantom{}_{\sim}}}
\def\sumint{\hbox{$\sum$}\!\!\!\!\!\!\!\,{\int}}

\let\Im\undefined
\let\Re\undefined
\DeclareMathOperator{\Re}{Re}
\DeclareMathOperator{\Im}{Im}

\DeclareMathOperator{\Tr}{Tr}

\def\gev{{\, \rm GeV}}
\def\mev{{\, \rm MeV}}

\usepackage{soul}

\def\Tr{{\rm Tr}}
\def\eps{\varepsilon}
\def\om{\omega}

\def\fs{\slashed}
\def\nn{\nonumber \\}

\usepackage[d]{esvect}

\def\XXint#1#2#3{{\setbox0=\hbox{$#1{#2#3}{\int}$}
		\vcenter{\hbox{$#2#3$}}\kern-.5\wd0}}

\usepackage{diagbox}

\def\sumint#1#2{\hbox{$\sum$}\!\!\!\!\!\!\!\,{\int\limits_{#1}^{#2}}}

\def\qq{\langle \bar q q \rangle}

\begin{document}
\title{Effect of mesonic off-shell correlations in the PNJL equation of state}

\author{K. Maslov}
\email{maslov@theor.mephi.ru}
\affiliation{AKFA University,
	Milliy bog street 264,
	111221 Tashkent, Uzbekistan}
\affiliation{Institute of Theoretical Physics, 
	University of Wroclaw, 
	Max Born place 9, 
	50-204 Wroclaw, Poland}

\author{D.~Blaschke}
\email{blaschke@ift.uni.wroc.pl}
\affiliation{Institute of Theoretical Physics,
	University of Wroclaw, 
	Max Born place 9, 
	50-204 Wroclaw, Poland}

\date{\today}
\begin{abstract}
	We study the meson contribution to the equation of state of the 2-flavor PNJL model, including the full momentum dependence of the meson polarization loops. Within the Beth-Uhlenbeck approach, we demonstrate that the contribution from the quark-antiquark continuum excitations in the spacelike region $\omega^2 - q^2 < 0$, i.e. the Landau damping, leads to an increase of the pressure for temperatures $\gsim 0.8\,T_c^\chi$ and a significant meson momentum cut-off dependence in the mesonic pressure and the QCD trace anomaly. We investigate the dependence of the results on the choice of the Polyakov-loop potential parameter $T_0$. From the dependence of the mesonic pressure on the current quark mass, by means of the Feynman-Hellmann theorem, we evaluate the contribution of the pion quasiparticle gas and Landau damping to the chiral condensate.
\end{abstract}
\maketitle

\section{Introduction}
Elucidating the structure of the phase diagram of quantum chromodynamics (QCD) is one of the challenging problems in particle physics. Its experimental exploration is presently performed at large existing heavy-ion collision (HIC) facilities, such as the Relativistic Heavy-Ion Collider (RHIC) at BNL Brookhaven, the Large Hadron Collider (LHC) and the Super Proton Synchrotron (SPS) at CERN Geneva. It will in future be continued with dedicated experiments at the Facility for Antiproton and Ion Research (FAIR) at GSI Darmstadt, the Nuclotron-based Ion Collider fAcility (NICA) at JINR Dubna and 
the Japan Proton Accelerator Research Complex (J-PARC) at the Tokai campus of JAEA. 
In nature, the QCD transition from a quark-gluon plasma (QGP) to a phase of confined quarks and gluons, a gas or liquid of hadronic resonances, is realized in the Early Universe and may also occur in the interiors of neutron stars, their mergers and supernova explosions, being subject to observations by multi-messenger Astronomy.

While at high energies QCD is in the regime of asymptotic freedom allowing to apply the well-developed methods of perturbation theory, 
a theoretical description of the hadronization transition with its aspects of dynamical chiral symmetry breaking and confinement (at low temperatures also color superconductivity) is a notoriously difficult task. It concerns the challenging domain of low energy QCD, where nonperturbative methods must be developed and applied.
A first principles approach to describe the QCD transition at finite temperatures uses Monte Carlo simulations of the QCD partition function formulated in lattice quantum field theory on a discretized Euclidean space-time. 
The $\mu_B = 0$ sector of the QCD phase diagram is well explored within lattice QCD, and a crossover transition with a pseudocritical temperature $T_c^{\rm LQCD} = 156.5 \pm 1.5$ MeV is obtained
\cite{HotQCD:2018pds}. 
This method, however, is inapplicable at large values of the baryon chemical potential $\mu_B$ because of the yet unsolved sign problem. 
At low temperatures and densities $n \simeq (1 - 2)\,n_0$, where $n_0\simeq 0.16\,{\rm fm}^{-3}$ is the nuclear saturation density, 
the chiral effective field theory approach for the description of 
hadronic degrees of freedom became a standard tool to describe 
nuclear phenomena with theoretically controlled uncertainties~\cite{Tews:2012fj}. However, these uncertainties become larger at larger densities, which renders it unsuitable for a description of dense matter, e.g. in neutron star interiors. Additional complications arise while extending this approach to large temperatures. 

With an increase of the density the QCD vacuum can change from a state with massive constituent quarks to one with light current quarks, referred to as the chiral symmetry restoration transition. Another phase transition relevant especially for astrophysical applications is the appearance of color superconductivity at large densities, which is related to the appearance of diquark condensates. This region of the QCD phase diagram is beyond the scope of the current work.

The first step in theoretically assessing the hadronization of quark and gluon degrees of freedom in a thermal quantum field theory is to 
define an effective Lagrangian in the quark sector of QCD that shares the essential symmetries with the QCD Lagrangian but is formulated in terms of quark currents that should emerge from "integrating out" the dynamical gluon degrees of freedom. When this current-current form of the interaction Lagrangian is symmetric under chiral rotation  of the quark fields one arrives at a version of the Nambu--Jona-Lasinio (NJL) model of low-energy QCD, see
\cite{Klevansky:1992qe,Ebert:1994mf,Hatsuda:1994pi,Buballa:2003qv}
for standard reviews. A modern refinement of the model takes into account the coupling of quark Dirac-spinors to a gluon background field in the Polyakov gauge
\cite{Meisinger:1995ih, Fukushima:2003fw, Megias:2004hj, Ratti:2005jh, Contrera:2007wu} which is called the Polyakov-loop improved NJL model or in short: PNJL model.
The traced Polyakov-loop $\Phi$ and its conjugate $\bar \Phi$ appear now as additional order parameters in the model which are vanishing in the confined phase due to the Z(3) center symmetry of SU(3) color and signal deconfinement when they approach unity under broken center symmetry. 
The thermodynamics of this PNJL model for the QGP is generally considered at the mean-field level, but the parameters of the model 
can be tuned so as to reproduce the pion mass $m_\pi=140$ MeV and pion decay constant $f_\pi=93$ MeV as well as the chiral light quark condensate $\langle \bar{u}u \rangle^{1/3} = - 240 \pm 20$ MeV~\cite{Buballa:2003qv}
(or a suitably chosen constituent quark mass) in the vacuum, which satisfies the low-energy QCD theorems,
the Gell-Mann--Oakes--Renner \cite{Gell-Mann:1968hlm} and the Goldberger-Treiman \cite{Goldberger:1958tr} relations.
To construct the pion simultaneously as the Goldstone boson  of the broken chiral symmetry and a pseudoscalar bound state in the pseudoscalar quark-antiquark channel in the vacuum, one quantizes the Gaussian fluctuations around the mean-field (MF) solution. 
This is achieved by solving the homogeneous Bethe-Salpeter equation, which defines the bound state pole of the pion propagator via the pseudoscalar-pseudoscalar polarization loop integral.
The NJL and PNJL model details can be found, e.g., in Ref.~\cite{Hansen:2006ee, Blaschke:2016sqn}.

Refined versions of the (P)NJL models include beyond-mean-field corrections according to the 1/$N_c$ power counting~\cite{Quack:1993ie, Hufner:1994ma, Blaschke:1995gr, Nikolov:1996jj}. 
The inclusion of such next-to-leading order terms requires particular care in order to keep the model consistent with the chiral symmetry. It was performed in both local and non-local formulations of the NJL and PNJL models and results in a change of the meson properties~\cite{Plant:2000ty, Oertel:1999fk, Oertel:2000sr}, chiral condensate dependence on the temperature~\cite{Oertel:2000sr,Radzhabov:2010dd}, quark propagator~\cite{Muller:2010am}, as well as the equation of state~\cite{Torres-Rincon:2017zbr}. 
Another important feature often included in such models is the $N_f$ dependence of the Polyakov-loop effective potential proposed in~\cite{Schaefer:2007pw}, which effectively results in a decrease of the $T_0$ parameter of the potential, which in a pure Yang-Mills theory has a meaning of the critical temperature of the deconfinement phase transition $T_0 = 0.27\gev$. 

The pion spectral features in PNJL-based models generally consist of a quasiparticle peak corresponding to a long-lived propagating pion mode, and the ``continuum'' part, which describes damped quark-antiquark correlations. At zero momentum in the frequency region $\om > 2\,m(T)$, where $m(T)$ is the constituent quark mass at temperature $T$, the pion decay into a $\bar q q$ pair is possible. This leads to a non-zero ``correlation'' contribution to the pion spectral function. With an increase of the temperature $m(T)$ rapidly decreases near the pseudocritical temperature of the chiral phase transition $T_c^\chi$ and the threshold for $\bar q q$ pair production becomes lower than the pion mass. It enables the pion decay into the $\bar q q$ pairs and therefore there are no more long-lived pionic excitations in the system, which is dubbed ``pion melting'' due to chiral symmetry restoration. 

For a finite momentum, the pion spectral function has a so-called Landau damping (LD) region with $\omega^2 < q^2$, corresponding to processes with absorption or emission of pions with the momentum $\vec q$ by the quark thermal bath~\cite{Weldon:1983jn,Hatsuda:1994pi}. This is a well-known feature observed in plasma physics, thermal QCD and QED~\cite{Baym:1992eu}, as well in the lattice simulations at finite momentum~\cite{Alberico:2006wc,Boguslavski:2018beu, Boguslavski:2021kdd}. Its effect on the chiral condensate within PNJL model was studied in~\cite{CamaraPereira:2020ipu}, but its contribution to thermodynamics was outside the scope of that work. In~\cite{Roessner_arxiv} the off-shell mesonic contribution to the pressure was included, but was absent in the published version of this paper~\cite{Roessner:2007gha}.
In many other implementations of chiral quark models the assumption of Lorentz-invariance of the pion polarization operator is often used, which allows obtaining the pion spectral function (phase shift) at finite momentum by simply boosting the $\vec q = 0$ result to a moving frame. As was shown in~\cite{Torres-Rincon:2017zbr}, the inclusion of the full momentum dependence changes the resulting dispersion relation of the pion quasiparticles very little in comparison to the vacuum case. Therefore, for the quasiparticle contribution, this assumption is well justified. Furthermore, for a given temperature the absolute value of the pion and sigma phase shifts in the LD region is exponentially suppressed at large momenta $q$~\cite{Yamazaki:2012ux,Yamazaki:2013yua}.
However, given that the Bose distribution function diverges as the frequency goes to zero, it is not possible to make a conclusion about the behavior of thermodynamic quantities compared to the free pion gas just from the qualitative consideration without further examining of the momentum and temperature dependence of this contribution.  

In this paper we demonstrate that the presence of the Landau cut in the meson propagators leads to a significant enhancement and a strong threshold dependence of the contribution to the total pressure from the pion and sigma correlations in this model. The work is based on the PNJL model in the simplest ``mean-field + fluctuations'' scheme without taking into account the meson contribution to the PNJL gap equation, similar to that used in~\cite{Ratti:2005jh,Blaschke:2016fdh,Blaschke:2016sqn,Yamazaki:2012ux,Yamazaki:2013yua}.

The work is organized as follows. In section II we describe the employed formalism of PNJL model and treatment of meson correlations. In section III we present our numerical results, focusing on the continuum contributions to thermodynamics from the meson correlations and their threshold dependence, and the section IV concludes the work. The explicit expressions for the meson polarization operator and an analytic estimate of the LD contribution to the pressure at low temperatures are given in the Appendix.


\section{Model setup}
\subsection{PNJL model}
We implement the PNJL model with $N_f = 2$ quark flavors and $N_c = 3$ colors ~\cite{Meisinger:2001cq, Fukushima:2003fw, Ratti:2005jh, Wergieluk:2012gd} at the baryon charge chemical potential $\mu$ and temperature $T$. The model is described by the Lagrangian
\begin{gather}
	{\cal L}_{\rm PNJL} = \bar q (i \fs D - m_0 ) q + G_s \Big[(\bar q q)^2 + (\bar q i \gamma^5 \vec \tau q)^2\Big],
\end{gather}
where $m_0 = 5.5 \mev$ is the bare quark mass assuming the isospin symmetry, $D_\mu = \partial^\mu - i \delta_0^\mu (A^0 + \mu)$ is the covariant derivative in the Polyakov-loop background field $A^0$ and the chemical potential $\mu$, and $\vec \tau$ are the the Pauli matrices in the isospin space. $G_s = 5.04\gev^{-2}$ is the NJL 4-quark coupling in the scalar-pseudoscalar channel.
Below the consideration will be limited to the $\mu = 0$ case, but we shall keep $\mu$ in the expressions for completeness. 

The traced Polyakov loop and its conjugate are defined as 
\begin{gather}
	\Phi(\vec x) = \frac{1}{N_c} \Tr_c \langle\langle L(\vec x) \rangle \rangle, \quad \bar \Phi(\vec x) = \frac{1}{N_c} \Tr_c \langle \langle L^\dagger(\vec x) \rangle \rangle\nn
	L(\vec x) = {\cal P} \exp \Big[i \int_0^\beta d \tau A_4 (\vec x, \tau)\Big],
\end{gather}
where $A_4 = i A_0$, and $\beta = 1/T$.
For a pure Yang-Mills (YM) SU(3) theory without quarks, we have $\bar\Phi = \Phi$, and the Polyakov loop is an order parameter for the deconfinement phase transition, corresponding to the spontaneous breaking of the Z(3) center symmetry. The thermodynamics of this phase transition in pure gauge theory for SU(3)$_c$ can be described well by the effective potential ${\cal U}(\Phi, \bar \Phi, T)$, which exhibits a unique minimum at $\Phi=0$ for low temperatures and develops a second minimum at $\Phi \neq 0$ as the temperature exceeds the critical temperature $T_0$, corresponding to a first-order SU(3) YM phase transition. In this work we use the same form as in~\cite{Ratti:2005jh} chosen to reproduce the SU(3) Yang-Mills thermodynamics obtained in lattice simulations~\cite{Boyd:1996bx} 
\begin{gather}
	\frac{{{\cal U}(\Phi, \bar \Phi, T)}}{T^4} = - \frac{b_2(T)}{2} \Phi \bar \Phi - \frac{b_3}{6} (\Phi^3 + \bar \Phi^3) + \frac{b_4}{4} (\bar \Phi \Phi)^2,\nn
	b_2(T) = a_0 + a_1\Big(\frac{T_0}{T}\Big) + a_2 \Big(\frac{T_0}{T}\Big)^2 + a_3 \Big(\frac{T_0}{T}\Big)^3,
\end{gather}
with the coefficients $a_{0, 1, 2, 3} = [6.75, -1.95, 2.625, -7.44]$, $b_3 = 0.75$, $b_4 = 7.5$. The parameter $T_0$ has the meaning of the critical temperature of the first-order phase transition in the pure YM theory and for this case $T_0 = 0.27\gev$. In presence of dynamical quarks the effect of running QCD coupling can be translated into a rescaling of this parameter, depending on the number of quark flavors~\cite{Schaefer:2007pw}.
In this paper we will compare our results between $T_0 = 0.27\gev$ and $T_0 = 0.208\gev$ obtained in~\cite{Schaefer:2007pw} for $N_f = 2$, and also for $T_0 = 0.178\gev$, corresponding to $N_f = 3$, in order to study the overall effect of such rescaling on the off-shell pion contribution to the EoS.

The total grand canonical thermodynamic potential in the PNJL model reads~\cite{Ratti:2005jh,Hansen:2006ee,Yamazaki:2012ux,Yamazaki:2013yua}
\begin{gather}
	\Omega(T; \Phi, \bar \Phi, m) = {\cal U}(\Phi, \bar \Phi, T) + \frac{(m - m_0)^2}{4 G_s} \nn - 2 N_f \Big\{\int_{|\vec p| < \Lambda }\hspace{-.8em}\frac{d^3 p}{(2 \pi)^3} N_c \eps_p + T\int\frac{d^3 p}{(2 \pi)^3} [z_\Phi^+ + z_\Phi^-] \Big\},
\end{gather}
where the coupling between quark and gauge sectors leads to a modification of the pressure integrand,
\begin{align}
	z_\Phi^+ & \equiv \Tr_c \ln[1 + L^\dagger y] = \ln \Big\{ 1 + 3(\bar \Phi + \Phi y)y + y^3 \Big\}, \\ z^-_\Phi & = z^+_\Phi(\Phi \leftrightarrow \bar \Phi, \mu \to - \mu), \quad y = e^{-(E_p - \mu)/T}, \nonumber
\end{align}
compared to the free quasi-fermion gas. For regularizing the vacuum contribution of the constituent quarks in this work we use the standard 3-momentum cutoff scheme, integrating only over quark momenta $|\vec p| < \Lambda$~\cite{Hatsuda:1994pi, Hansen:2006ee}.  We use $\Lambda = 651\mev$ throughout this work.

The convergent contribution of the thermal quark and antiquark excitations is integrated over all momenta. The mean-field (MF) equilibrium values of $\Phi, \bar \Phi$, and $m$ follow from minimization of the thermodynamic potential:
\begin{gather}
	\frac{\partial \Omega}{\partial \Phi} = 	\frac{\partial \Omega}{\partial \bar \Phi} = \frac{\partial \Omega}{\partial m} = 0.
\end{gather}

The last of these equation leads to the PNJL gap equation

\begin{gather}
	m - m_0 = 2 G_s N_f N_c \times 
	\nn
	\int\frac{d^3 p}{(2\pi)^3} \frac{2 m}{\eps_p} \Big[\theta(\Lambda - p) -  \Big(f_\Phi^+(E_p) + f_\Phi^-(E_p)\Big)\Big], 
	\label{eq::gap}
\end{gather}
where, using $y = e^{-\beta(\eps - \mu)}$,
\begin{align}
	f_\Phi^+(\eps) &= \frac{(\bar \Phi + 2 \Phi y) y + y^3}{1 + 3(\bar \Phi + \Phi y) y + y^3},	\label{eq::f_phi}
	\\ \nonumber f_\Phi^-(\eps) &= f_\Phi^+ (\eps; \Phi \leftrightarrow \bar \Phi, \mu \to -\mu).
\end{align}
In the limit $\Phi \to 1, \bar \Phi \to 1$, corresponding to the deconfined state, $f_\Phi^\pm$.

The numerical solution of the MF approximation is discussed in Section~\ref{sect3::MF}.

\subsection{Mesonic correlations}

The $\pi$- and $\sigma$-meson 1PI polarization operators read
\begin{gather}
	\Pi_M(\omega, \vec q) = T \sumint{P}{} G(\om_n, \vec p) \Gamma_M G(\om_n - \eps_k, \vec p - \vec q) \Gamma_M,
\end{gather}
where $M = \{\pi, \sigma\}$, $\Gamma_\pi = i\gamma^5, \Gamma_\sigma = 1$.
The imaginary part of the polarization loop contains no divergences and can be evaluated without regularization. Then we reconstruct the real part using the Kramers-Kronig relation. The explicit expression for $\Im\Pi_M$ is given in the Appendix~\ref{app::meson_exprs}.
The RPA-resummed propagator of a quasi-meson $M$ is then determined as 
\begin{gather}
	D_M(\om,  q) = - \frac{2 G_s}{1 + 2 G_s \Pi_M(\om,  q)}.
	\label{eq::propagator}
\end{gather}

We define the mass $m_M$ of a quasi-meson $M$ as a solution of 
\begin{gather}
	1 + 2 G_s \Re \Pi_M(\om = m_M, q=0) = 0.
	\label{eq::m_pi}
\end{gather}
The usefulness of this expression is limited to the temperature region in which the meson is a bound state. All the physical information about mesonic excitations in the medium is encoded in the spectral function
\begin{gather}
	\rho_M(\om, q) = - 2 \Im D_M(\om, q),
\end{gather}
which is the subject of our study in this work.

\subsection{Pressure of the meson gas}
\label{sect:PT}
After integrating out the quarks in the Gaussian approximation and dropping the vacuum mesonic pressure, the meson contribution to the pressure reads~\cite{Hufner:1994ma,Roessner_arxiv,Blaschke:2013zaa,Torres-Rincon:2017zbr}
\begin{gather}
	P_M = d_M \sum_{k = \rm QP, LD} \int_{|\vec q| < \Lambda^k}\frac{d^3 q}{(2\pi)^3}  w_M^{k}(q, T), 	\label{eq::fp} \\
	w_M^{\rm QP} \equiv \int\limits_{q}^\infty \frac{d \omega}{\pi} \frac{\delta_M(\omega, q, T)}{e^{\omega/T} - 1},\,\,
	w_M^{\rm LD} \equiv \int\limits_{0}^q \frac{d \omega}{\pi} \frac{\delta_M(\omega, q, T)}{e^{\omega/T} - 1},\nonumber
\end{gather}
where $d_\pi = 3, d_\sigma=1$, and 
\begin{gather}
	\delta_M(\omega, q, T) = - \arctan \frac{\Im D_M(\om, q, T)}{\Re D_M(\om ,q, T)}
	\label{eq::delta_pi}
\end{gather}
is the quark-antiquark scattering phase shift in the corresponding meson channel $b$.
We have separated the part coming from the Landau damping (LD) region $0 < \om < q$ from the rest, which we label as the quasipole (QP) contribution, with their respective momentum thresholds $\Lambda^{\rm QP}, \Lambda^{\rm LD}$.
For simplicity, we use $\Lambda^{\rm QP} \to \infty$, which gives a maximum possible quasipole contribution to the thermodynamics, that is weakly dependent on the choice of the threshold for $T$ below $T_c^\chi$. For the LD contribution in this work we will vary the momentum threshold $\Lambda^{\rm LD}$ in the limits $\Lambda^{\rm LD} = (1-2)\Lambda$ encountered in the literature~\cite{Nikolov:1996jj,Roessner_arxiv,Yamazaki:2012ux} in order to investigate its threshold sensitivity.

\section{Numerical results}
\label{sect3}
In this section we discuss the numerical results of the developed approach to mesonic correlations in the PNJL model. We start in subsect.~\ref{sect3::MF} with the MF approximation and proceed to discussing the meson spectral functions in subsect.~\ref{sect3::spectral} and their contribution to thermodynamic properties of PNJL matter in subsect.~\ref{sect::thermo}. Then we show the correction to the chiral condensate by means of the Hellmann-Feynman theorem arising from the LD contribution to the pressure in subsect.~\ref{sect::qq}. Finally, we discuss in subsect.~\ref{sect::varT0} the dependence of the results on rescaling the critical temperature parameter $T_0$ of the Polyakov loop potential for example of the interaction measure.


\subsection{Mean-field solutions and meson masses}
\label{sect3::MF}
We start with discussing the temperature dependence of the mean fields and meson masses in the PNJL model, as these quantities will later define the meson contribution to the pressure.
\begin{figure}
	\includegraphics[width=.9\linewidth]{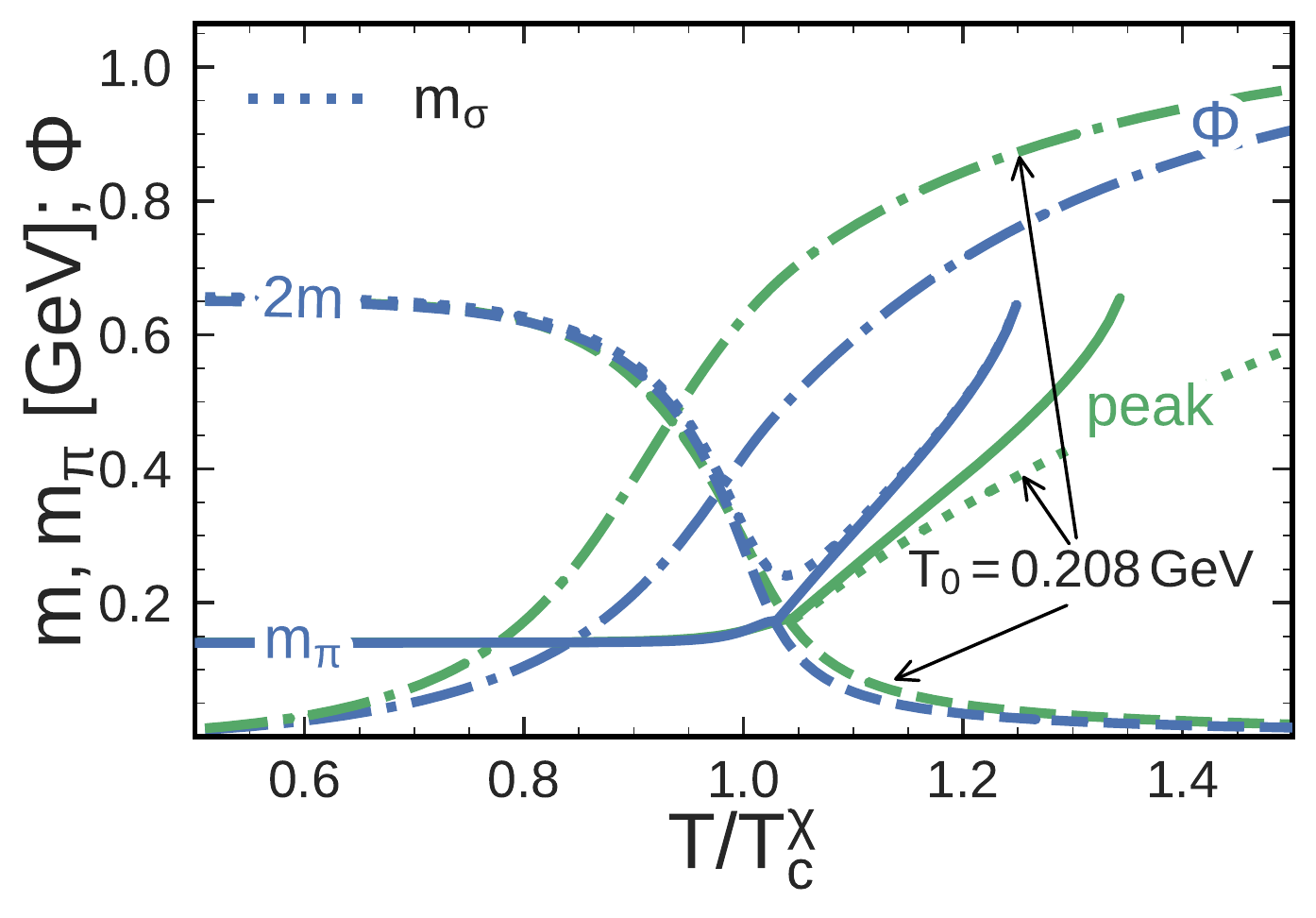}
	\caption{Masses of pions (solid lines) and constituent quarks (dashed lines), together with the Polyakov loop variable $\Phi$ (dash-dotted lines) as functions of the temperature for PNJL models with $T_0 = [0.27, 0.208]\gev$. The mass of the sigma is shown by the dotted line only for $T_0 = 0.27\gev$ case for clarity. For $T_0 = 0.208\, T_c^\chi$ we also show the position of the pion spectral function peak (dotted line) at $T > T_{\rm Mott}$.}
	\label{fig::MT}
\end{figure}
In Fig.~\ref{fig::MT} we compare the temperature dependence of the pion and sigma masses, constituent quark mass, and the Polyakov loop variable between two choices of the $T_0$ parameter of the Polyakov loop potential $T_0 = [0.27, 0.208]\gev$. 
The characteristic temperatures of the hadron-quark phase transition are the pseudocritical temperatures related to the chiral symmetry restoration $T_c^{\chi}$ and modeling of the deconfinement $T_c^{\Phi}$ , respectively, defined as
\begin{gather}
	T_c^\chi = \arg \max \frac{\partial \qq(T)}{\partial T}, \\  T_c^\Phi = \arg \max \frac{\partial \Phi(T)}{\partial T}.
\end{gather}
In Fig.~\ref{fig::MT} we plot the results as functions of $T$ normalized by the respective $T_c^\chi$ for these cases. The key parameters extracted from these solutions are collected in Table~\ref{tab::Tcs}.



\begin{table*}
		\begin{tabular}{|c|c|c|c|c|c|c|c|c|}
			\hline
			\diagbox{Model}{Quantity}& $T_c^\chi$ & $T_c^\Phi$ & $T_{\rm Mott}$ & \makecell[bc]{  $T_{c, \rm fl}^{\chi}$ \\ $ \Lambda _{\rm LD}=\Lambda$} & \makecell[bc]{  $T_{c, \rm fl}^{\chi}$ \\ $ \Lambda _{\rm LD}=2\Lambda$} & 
			$T_{\max}$ & 
			\makecell[bc]{  $T_{\max, \rm fl}$ \\ $ \Lambda_{\rm LD}=\Lambda$}\\\hline\hline
			$T_0 = 0.27\gev$ & 0.229 & 0.225 & 0.236 & 0.227 & 0.223 & 0.282 & 0.226 \\ \hline
			$T_0 = 0.208\gev$ & 0.201 & 0.183 & 0.209 & 0.196 & 0.192 & 0.204 & 0.198  \\ \hline
			$T_0 = 0.178\gev$ & 0.194 & 0.166 & 0.199 & 0.187 & 0.183 & 0.190 & 0.184 \\
			\hline
		\end{tabular}
		\caption{Comparison of characteristic temperatures in units of\gev\, between PNJL models with various $T_0$. $T_c^\chi$ and $T_c^\Phi$ are the pseudocritical temperatures of the chiral and Polyakov-loop phase transition temperatures, $T_{c, \rm fl}^\chi$ are the $T_c^\chi$ recalculated with inclusion of the pion contribution as described in the section~\ref{sect::qq}, and $T_{\rm max}$, $T_{\max,\rm fl}$ are the locations of the maximum of the trace anomaly $I(T)/T^4$, see section~\ref{sect::varT0}.}
		\label{tab::Tcs}
	\end{table*}

\begin{figure*}[!]
	\includegraphics[width=.95\linewidth]{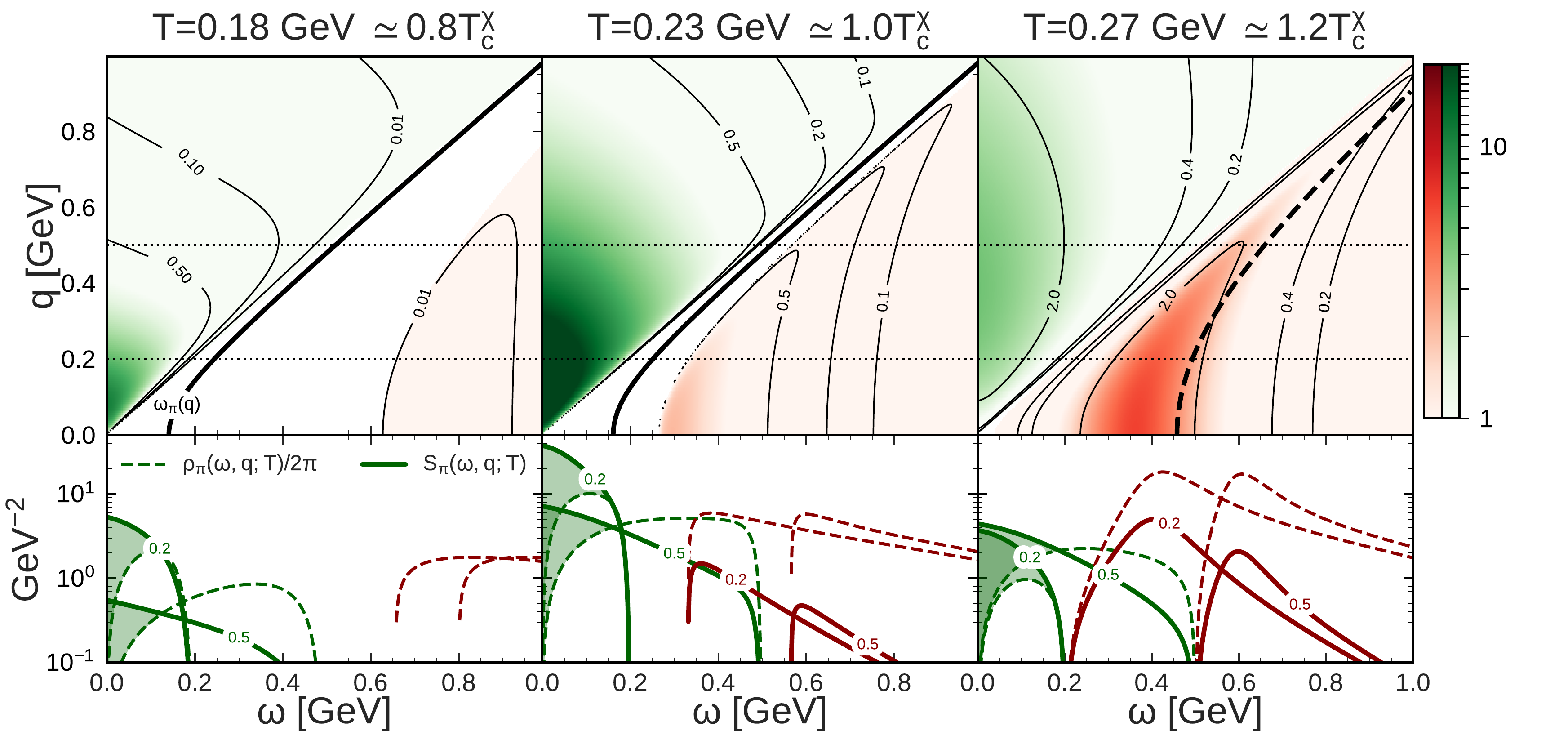}
	\caption{Upper panels: dynamic structure factor $S_\pi(\om, q; T)$ in the pion channel in the units of $\gev^{-2}$ in the $\omega - q$ plane for temperatures $T = [0.18, 0.23, 0.27] \gev \simeq [0.8, 1.0, 1.2]\,T_c$. Thick lines show the position of the pion pole for $T = 0.18\gev$ and $0.23\gev$ (solid lines) and the solution of the dispersion relation for $T = 0.27\gev$, at which pion is dissociated (dashed line). Lower panels: the regular part of the pion spectral function $\rho_\pi(\om, q, T)$ and the dynamics structure factor $S_\pi(\om, q; T)$ as functions of the frequency for two momenta $q = [0.2, 0.5] \gev$ indicated on the lines and shown by the dashed horizontal lines in the upper panels. Shaded areas in the lower panels illustrate the enhancement of the spectral function due to the thermal factor.}
	\label{fig::rho_pi_heat}
\end{figure*}

The temperature behavior of $m$ and $\Phi$ is standard for the PNJL model~\cite{Hansen:2006ee, Yamazaki:2012ux}. With increasing temperature the quark mass monotonously decreases, while the Polyakov loop expectation value grows. The results for $m(T/T_c^\chi)$ are almost coinciding between the $T_0 = [0.27, 0.208]\gev$ cases, with the respective $T_c^\chi$. The main difference between the cases $T_0 = [0.27, 0.208]\gev$ is the well-known increase of the difference between $T_c^\Phi$ and $T_c^\chi$~\cite{Ratti:2005jh, Digal:2000ar}. Therefore, for $T_0 = 0.208\gev$ the $\Phi(T)$ and, correspondingly, the quark distribution function~\eqref{eq::gap} is larger even for low $T/T_c^\chi$ than in the $T_0 = 0.27\gev$ case.

The pion and sigma masses are defined as solutions of Eq.~\eqref{eq::m_pi}. The pion pole position stays almost constant below the corresponding critical temperature. For $T \lsim T_c^\chi$ the $\sigma$ mass is always slightly above the $2m$ threshold and the $\sigma$ propagator has a quasipole with a finite quasiparticle width. For temperatures exceeding the pion Mott temperature $T_{\rm Mott}$, defined by $m_\pi(T_{\rm Mott}) = 2 m (T_{\rm Mott})$, the pion quasiparticles acquire finite width related to the decays into quark-antiquark pairs, as well as a thermal mass, which rises approximately linearly in temperature. The $\sigma$ quasipole location follows the threshold mass $2 m(T)$ up to $T \simeq T_c^\chi$ and at larger temperatures coincides with the thermal mass of the pion, reflecting the chiral symmetry restoration. 
For $T \gsim [0.29, 0.27] \gev$ in respective cases $T_0 = [0.27, 0.208]\gev$, Eq.~\eqref{eq::m_pi} does not have a real solution at all. Therefore, the lines for $m_{\pi, \sigma}(T)$ in Fig.~\ref{fig::MT} terminate at these temperatures in units of respective $T_c^\chi$.
The peak position of the spectral functions, shown for the $T_0 = 0.208\gev$ case as an example, stays linearly increasing for all $T > T_c^\chi$. The peak position is always below the real solution of the dispersion equation~\eqref{eq::m_pi}, as was also observed in~\cite{Hansen:2006ee}.

As was mentioned above, the main effect of decreasing $T_0$ on the mean fields is the increase of the gap between $T_c^\Phi$ and $T_c^\chi$. More subtle differences are the slight decrease of the quark mass below $T_c^\chi$ for the case $T_0 = 0.208\gev$, and the noticeable decrease of the slope of the pion thermal mass at $T/T_c^\chi > 1$ with respective $T_c^\chi$. In terms of the absolute $T$ the latter effect is still present, but less noticeable in the magnitude. These observations will prove to be useful for comparing the contribution of the low-energy pionic excitations to the thermodynamics between the cases $T_0 = [0.27, 0.208]\gev$ in the section~\ref{sect::varT0}.


\subsection{Pion spectral properties}
\label{sect3::spectral}
In order to intuitively compare the contribution of different regions of the spectral strength in the pion channel to thermodynamic quantities, it is useful to consider the dynamic structure factor in the axial channel,
\begin{gather}
	S_\pi(\om, q; T) =  \frac{1}{2\pi} \frac{\rho_\pi(\om, q; T)}{e^{\om/T} - 1},
	\label{eq::rho_T}
\end{gather}
as it was used, e.g., in ~\cite{Nishimura:2022mku} to illustrate the dilepton yield enhancement due to the diquark dynamic fluctuations. For $\om < q$, where the bound state is absent, its behavior is qualitatively the same as of the quantity $\frac 1 \pi \delta_\pi/(e^{\om/T} - 1)$ entering Eq.~\eqref{eq::fp}. Therefore the plot of $S_\pi(\om, q; T)$ allows to illustrate the enhancement of pressure integrand for $\om \to 0$ and suppression at $\om \gg T$.

In Fig.~\ref{fig::rho_pi_heat} we show $S_\pi(\om, q; T)$ for temperatures $T = [0.18, 0.23, 0.27]\gev \simeq [0.8, 1.0, 1.2]\,T_c^\chi$ in the PNJL model with $T_0 = 0.27\gev$. The upper panels show the heatmap plot of this quantity in the $\om - q$ plane. For $T = [0.8, 1.0]\,T_c^\chi $ the thick solid lines represent the poles of the pion propagator~\eqref{eq::propagator} present the real axis. For $T = 1.2\,T_c^\chi$ the pion is dissolved, and the dashed line in the corresponding panel shows the zero of the real part of the denominator of~\eqref{eq::propagator}. The shown contour levels are adjusted to be the same in the pair production region and the LD region of the spectral function.
The main feature of this plot is the enhancement of the LD region of $S_\pi(\om, q; T)$ by the thermal factor even at $T = 0.8\,T_c^\chi$, where the quark and pion masses are close to their respective vacuum values. However, for $T = [0.8, 1.0]\,T_c^\chi$ the contribution to the pion pressure is governed mainly by the pole contribution, see also Fig.~\ref{fig::fp} and the description in the text. For $T = 1.2\,T_c^\chi$ the threshold mass $2 m \simeq 0.1\gev$ is below the pion mass, and the pion pole is absorbed into the continuum. In this case it is clearly seen that for $q \gsim 0.5\gev$ the LD region gives the dominant contribution to the weighed spectral function. 

In the lower panels of the Fig.~\ref{fig::rho_pi_heat} we show $\rho_\pi(\om, q; T)$ and $S(\om, q; T)$ for momenta $q = [0.2, 0.5]\gev$ indicated on the lines. We see that the presence of the thermal factor enhances the spectral function in the low-frequency region by an order of magnitude, which is shown in the figure by the shaded areas. Moreover, in the limit $\om \to 0$ $S_\pi(\om, q;T) \to {\rm const}$ contrary to $\rho_\pi(\omega, q; T) \to 0$ being an odd function of the frequency. Simultaneously, the pair production region $\om^2 > 4 m^{2} + q^2$ is suppressed by the thermal factor. For $T \simeq 1.2\,T_c$ the weighted spectral function $S_\pi(\om, q; T)$ in the LD region exceeds the quasipole spectral function for $q = 0.5\gev$.


\begin{figure}
	\centering
	\includegraphics[width=.9\linewidth]{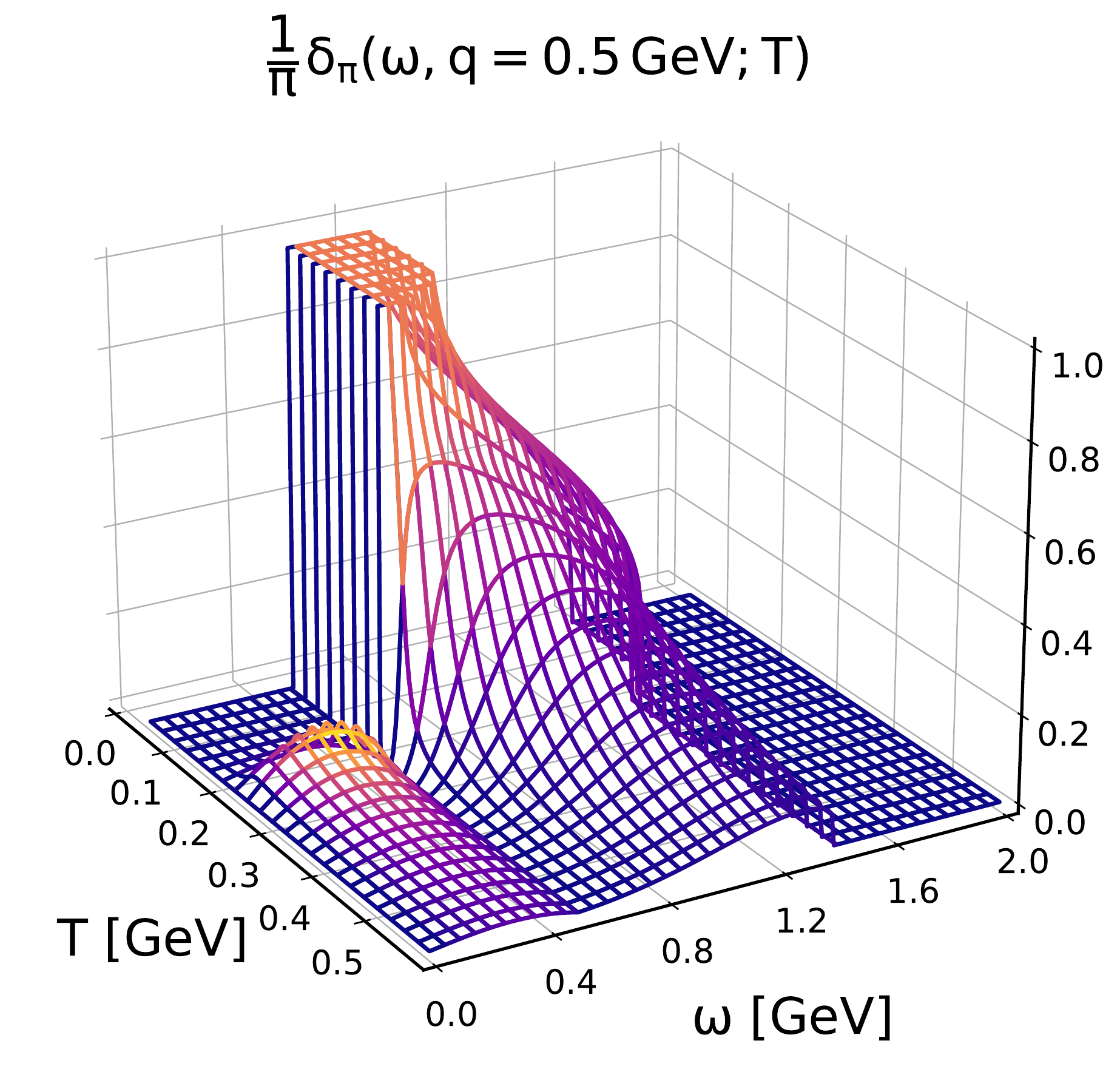}
	
	\includegraphics[width=.9\linewidth]{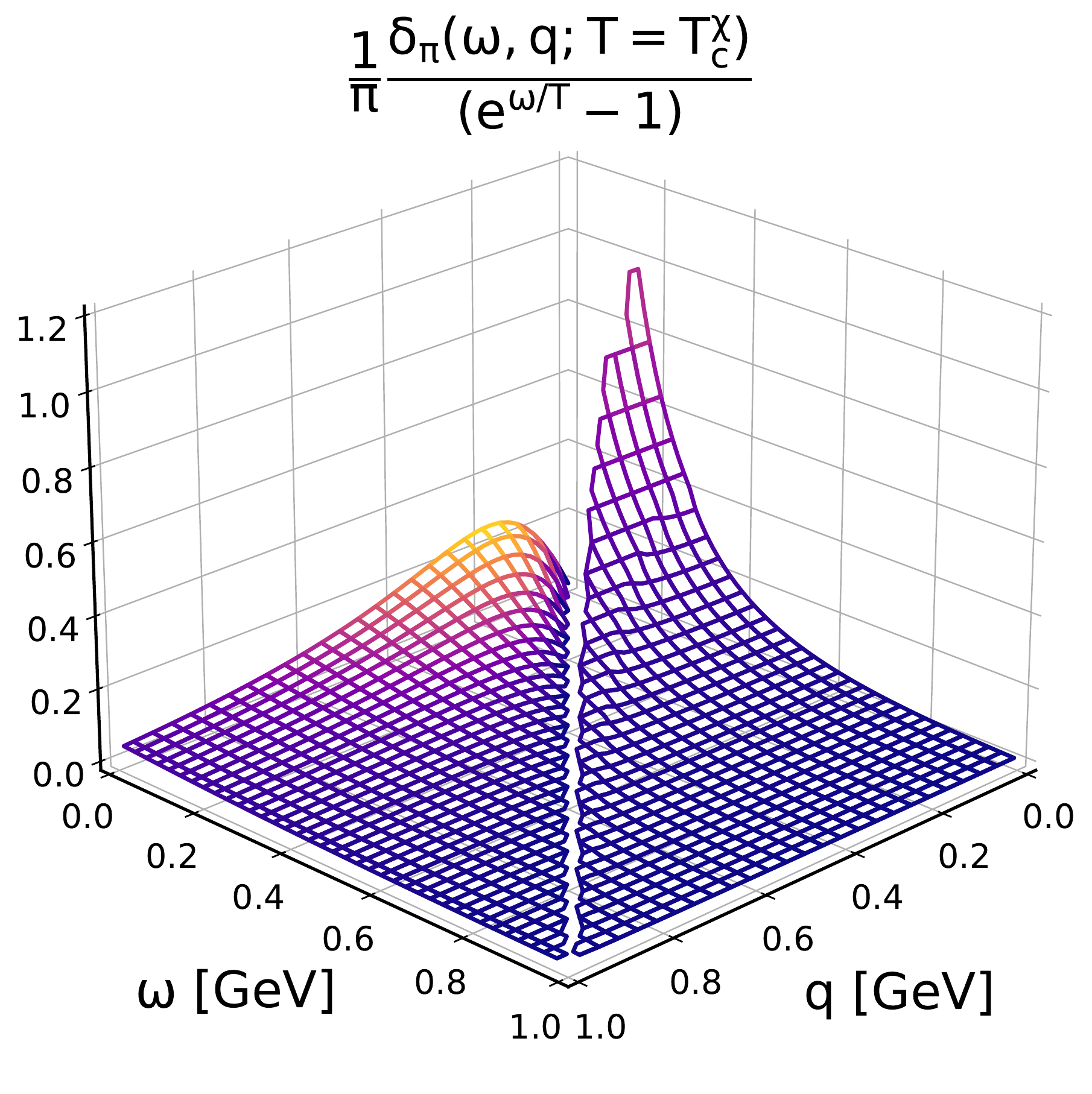}
	\caption{Upper panel: Pion phase shift in the PNJL model with $T_0 = 0.27\gev$ as a function of the frequency and temperature for a fixed momentum $q = 0.5\gev$. Lower panel: Pion phase shift weighed with the Bose factor as a function of momentum and frequency for $T = T_c^\chi$. Separate color mappings are used for $\om < q$ and $\om > q$ in both panels.}
	\label{fig::delta_3d}
\end{figure}


In the upper panel of Fig.~\ref{fig::delta_3d} we show the pion phase shift as a function of the frequency and temperature for a fixed momentum $q = 0.5\gev$. For the temperatures $T < T_{\rm Mott}$ the phase shift shows a jump from $0$ to $\pi$ at $\omega \simeq \sqrt{m_\pi^2(T) + q^2}$ corresponding to the existence of pion as a bound state. The LD contribution is negligibly small for $T \lsim 0.2 \gev$.
As the temperature increases, for $T \gsim T_{\rm Mott}$ the pion bound state is dissolved, so the phase shift never reaches $\pi$ and changes continuously for $\omega > 2 m(T)$, which corresponds to the resonance-shape structure in $\rho(\om, q;T)$ in the lower right panel of Fig.~\ref{fig::rho_pi_heat}. Simultaneously the LD contribution becomes noticeable, with a characteristic maximum at $\omega \simeq q/2$ and  $T \simeq T_c^\chi$. The magnitude of the LD contribution is still around 5 times smaller than the quasipole part for $T \simeq T_c^\chi$. For $T \gsim T_c^\chi$ the LD part of the phase shift decreases with an increase of the temperature, which is due to the growth of the pion thermal mass characterizing the overall magnitude of $\Re \Pi_\pi(\omega, q; T)$ that enters the denominator in $\eqref{eq::delta_pi}$.

In the lower panel of Fig.~\ref{fig::delta_3d} we show the integrand of the mesonic contribution to the pressure in Eq.~\eqref{eq::fp}, being the pion phase shift multiplied by the Bose factor, at the critical temperature $T = T_c^\chi = 0.229\gev$. In the QP region $\omega > q$ the thermal factor leads to an exponential suppression of the integrand at large $\omega$. In the LD region, as a consequence of thermal enhancement, the integrand becomes comparable with the one in the QP region for $q \gsim 0.5\gev$ and exceeding it at large momenta. The reason for this is that different characteristic temperatures govern the large-momentum tails of the corresponding contributions, being $\exp({-\sqrt{(2m)^2 + q^2}/2 T})$ for the LD part and $\exp({- \sqrt{m_\pi^2(T) + q^2}/T})$ for the QP contribution, which can be seen from the explicit expressions in the Appendix~\ref{app::analytic}.

\subsection{Mean-field + fluctuations thermodynamics}
\label{sect::thermo}
\begin{figure*}
	\centering
	\includegraphics[width=\linewidth]{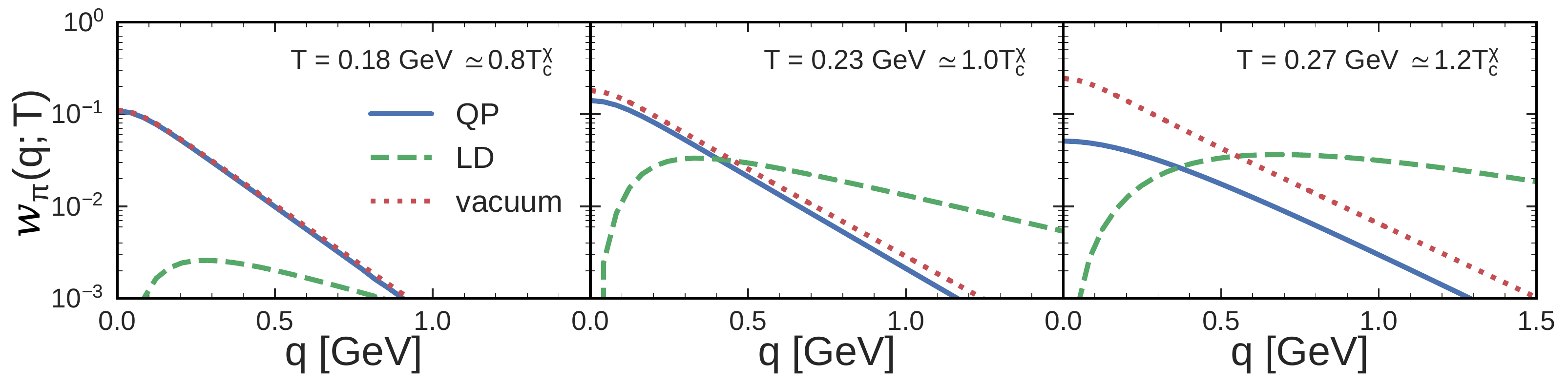}
	\caption{The QP (solid lines) and LD (dashed lines) contributions to the pressure integrand $w_\pi(q; T)$ \eqref{eq::fp} as functions of momentum for temperatures $T = [0.18, 0.23, 0.27] \gev \simeq [0.8, 1.0, 1.2]\,T_c^\chi$. Dotted lines show the free pion gas case.}
	\label{fig::fp}
\end{figure*}


Let us start with examining the contribution of the regions of the spectral density described above to the mesonic pressure integrand $w_\pi(q;T)$, defined in~\eqref{eq::fp}.
In Fig.~\ref{fig::fp} we plot $w_\pi(q; T)$ as a function of pion momentum for the temperatures $[0.18, 0.23, 0.27]\gev \simeq [0.8, 1.0, 1.2]\,T_c^\chi$, where for the Polyakov loop potential the $T_0 = 0.27\gev$ is chosen. For $T = 0.18\gev$ this function almost coincides with the free pion gas case
\begin{gather}
	w_\pi^{\rm vac}(q; T) = -T \ln (1 - e^{-\sqrt{m_\pi^2(T=0) + q^2}/T}),
\end{gather}
because the $\bar q q$-pair continuum threshold is still large and the pair production region, which gives a negative contribution to $w^{\rm QP}$, is exponentially suppressed by the thermal weight. The LD contribution is two orders of magnitude smaller than the QP one. With an increase of the temperature to $0.2 \gev$ the $\bar q q$-pair threshold becomes lower, which enhances the negative contribution of the $\omega>2m$ region to the pion distribution function. For $T = 0.27\gev \simeq 1.2\, T_c^\chi$ the pion pole moves away from the real axis and the distribution function is suppressed compared to the vacuum case. This corresponds to the Mott dissociation of pions and leads to a decrease of the pion QP contribution to the pressure for $T \gsim T_c^\chi$, which is well-described in the literature~\cite{Blaschke:2013zaa, Yamazaki:2012ux}.

However, simultaneously with an increase of the temperature, the LD contribution starts playing a noticeable role. The large-momentum tail of $\Im\Pi_{\rm LD}$ involves $2\,T$ in the exponent, cf. Eqs. \eqref{eq::J_LD_0}, \eqref{eq::J_LD_ollq}, and is therefore enhanced in comparison with the free pion gas case, as can be seen from the middle and right panels of the Fig.~\ref{fig::fp}. For $T \simeq [1, 1.2] \, T_c$ the LD contribution to $w_\pi(q; T)$ is at least comparable to the pole one, due to both its large value in the maximum and twice the effective temperature in its large-momentum tail. However, for low temperatures this expression is suppressed by the same Polyakov-loop mechanism as quark contribution to the thermodynamics,  despite that the Polyakov loop acts only on the colored partons. It follows from the same expression~\eqref{eq::J_LD_ollq}, which shows that $\Im\Pi_{\rm LD}$ is proportional to the quark distribution function.

\begin{figure}
	\centering
	\includegraphics[width=.95\linewidth]{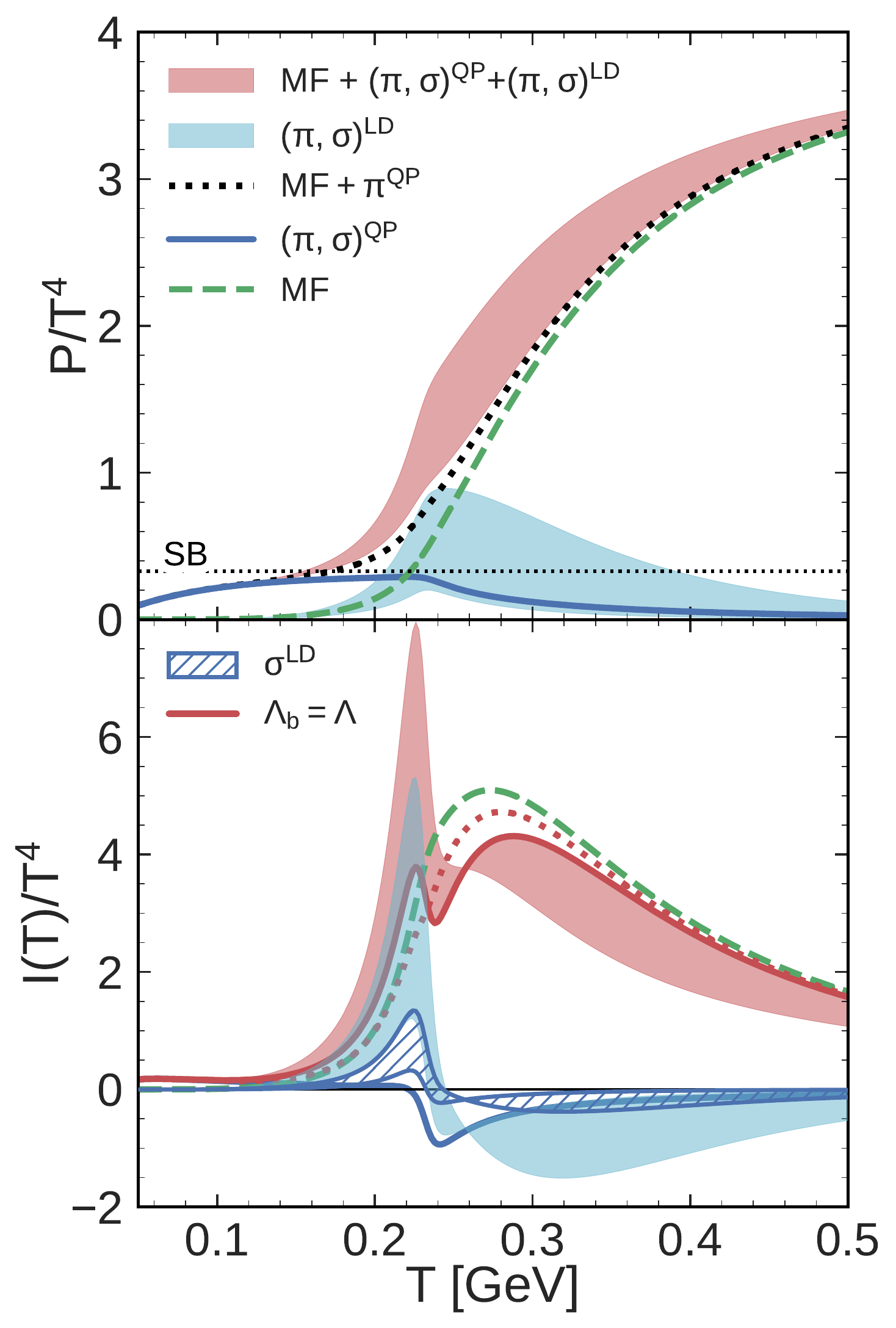}
	\caption{Upper panel: pressure as a function of the temperature in the PNJL model with $T_0 = 0.27\,\gev$ with the contribution from the meson correlations. Solid line denotes the QP contribution of the excitations with $\om > q$, dashed line stands for the PNJL pressure in the MF approximation, and dotted line is the sum of MF and QP pressure. Shaded areas show the uncertainty in the LD contribution and the total pressure coming from varying the meson threshold $\Lambda^{\rm LD}$ in the limits $[1 - 2]\,\Lambda$. Lower panel: same contributions to the trace anomaly as a function of the temperature. Hatched area is the LD contribution from the $\sigma$-meson. The $\Lambda^{\rm LD} = \Lambda$ case for the total $I(T)$ is highlighted by a solid line.}
	\label{fig::pres}
\end{figure}

The correlation pressure then follows from integrating the $w_{\pi}^{\rm QP, LD} (q, T)$ over the momentum. In the upper panel of Fig.~\ref{fig::pres} we show the contributions to the pressure coming from the MF approximation to the PNJL model and the meson fluctuations, with and without inclusion of the LD region for $T_0 = 0.27 \gev$.  Without the LD contribution meson pressure (solid line) grows with an increase of the temperature towards the Stephan-Boltzmann (SB) limit (dotted line) below $T_{\rm Mott} \simeq T_c^\chi$ along the result for the ideal pion gas with the vacuum pion mass (dash-dotted line). For $T > T_{\rm Mott}$ the pion pressure decreases as the pion bound state dissolves due to the lowering of $\bar q q$ continuum threshold and the increase of the pion thermal mass. This is the usual observed behavior of the pressure with Mott dissociation of the bound states~\cite{Roessner:2007gha,Yamazaki:2012ux,Blaschke:2013zaa}.

However, with the increase of the temperature the LD contribution starts to play a noticeable role. We plot it as a band corresponding to varying the LD cutoff $\Lambda^{\rm LD}$ from $\Lambda$ to $2 \Lambda$ in order to demonstrate its sensitivity. The temperature dependence of the contribution to the pressure from the LD region at $T\lsim T_c^\chi$ is qualitatively in the same as the MF pressure, as it is suppressed by the same thermal exponential factor, see Appendix~\ref{app::analytic}. An analytic estimate for the essential threshold dependence of $P_\pi^{\rm LD}$ is given in Appendix~\ref{app::analytic}. As the temperature increases, for any given $\Lambda^{\rm LD}$ this contribution to the pressure has a maximum near $T_c^\chi$ and decreases with further increase of the temperature due to the presence of the momentum threshold.
The overall result for the LD excitation pressure is heavily dependent on the  of the the LD threshold $\Lambda_\pi$, because the large high-momentum tail of the pressure integrand, as shown in Fig.~\ref{fig::fp}. We see that the cutoff-related uncertainty can reach $2\,P_{\rm SB}^{\pi}$ for $T \simeq T_c^\chi$.

Another quantity characterizing the thermodynamics of the model is the scaled trace anomaly $I(T) = T^5 d(P/T^4) / dT$ shown in the lower panel of Fig.~\ref{fig::pres}, which is more sensitive to various contributions to the pressure. The monotonous increase in the PNJL pressure results in a positive contribution to $I(T)$, which has a characteristic maximum. The quasipole correlation contribution is positive for $T \lsim 0.8 \, T_c$ and becomes negative as the temperature increases, since the quasipole meson pressure decreases as the mesons melt and acquire the thermal mass, leading to the decrease of the QP pressure. The LD contribution to $I(T)$ also changes sign as the presence of the LD threshold leads to decreasing LD pressure with an increase of the temperature. We note that even for a conservative choice of $\Lambda^{\rm LD} = \Lambda$ the contribution of the LD spectral region to $I(T)$ is of the same order as the QP one. For $T \gsim T_c$ they both are negative and each lead to a significant decrease of the overall $I(T)$.

The presence of the kink near $T_c$ in the overall $I(T)$ is related to the rapid growth of the LD contribution as $T$ approaches $T_c^\chi$ from below. For a larger threshold $\Lambda^{\rm LD} = 2\Lambda$ the kink is eliminated, since the LD contribution is large enough to overwhelm the second maximum. Therefore, in such a model the position of the characteristic maximum of $I(T)$ is affected by the magnitude of the mesonic LD contribution. Later in the subsect.~\ref{sect::varT0} we will see that even for a fixed threshold lowering of the $T_0$ parameter leads to an enhancement of the LD contribution and a similar effect on the position of the $I(T)$ maximum.

\subsection{Perturbative correction to the chiral condensate}
\label{sect::qq}

As it is seen from Fig.~\ref{fig::pres}, in PNJL model at temperatures below $T_c$ the dominant contribution to the pressure comes from the gas of mesonic excitations and the quark degrees of freedom are statistically suppressed by the Polyakov loop. The low-$T$ behavior of the chiral condensate in a physically acceptable model should be governed by the pion contribution. 

The pion polarization operator itself depends on the constituent quark mass, which allows us to calculate the ``perturbative'' contribution to the chiral condensate from the meson $M = (\pi,\sigma)$ fluctuation gas as
\begin{gather}
	\langle \bar q q \rangle_M = - \frac{\partial P_M(T; m)}{\partial m_0}.
\end{gather}

The presence of mesonic fluctuations does affect the chiral condensate In the model ``mean field + meson fluctuations'' that we use in this study the back-relation of the meson excitations to the quark MF sector is not included. A way of estimating the pion contribution to the chiral condensate perturbatively is to take the derivative over the quark current mass $m_0$ at the MF solution for $m$:
\begin{gather}
	\langle \bar q q \rangle_{\rm tot} = \langle\bar q q \rangle_{\rm MF} + \langle\bar q q \rangle_{\pi, \sigma}^{\rm QP} + \langle\bar q q \rangle_{\pi, \sigma}^{\rm LD}.
\end{gather}

\begin{figure}
	\centering
	\includegraphics[width=.9\linewidth]{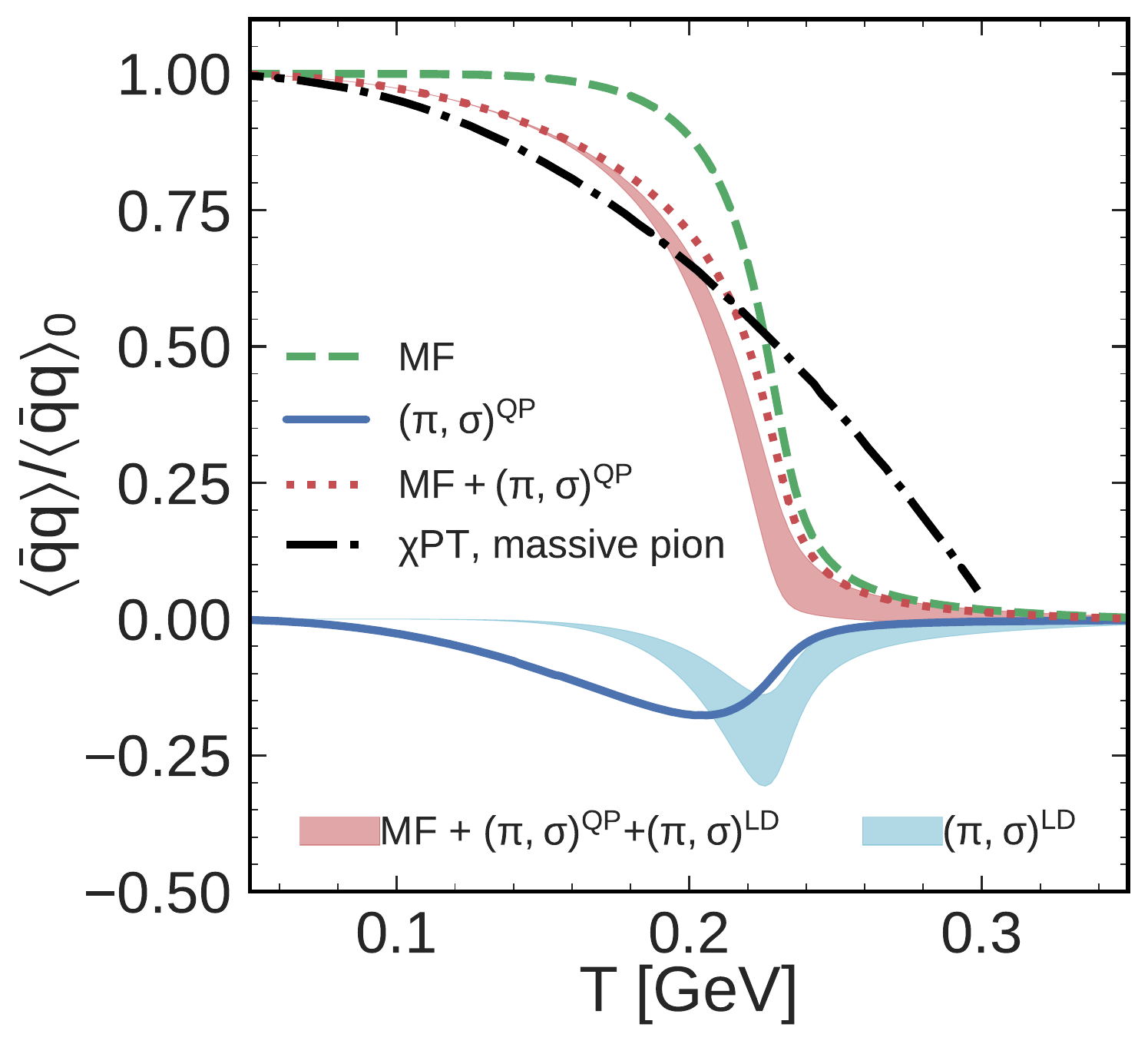}
	\caption{Contributions to the chiral condensate as functions of the temperature for $\Lambda_T = \Lambda$.  The line and shaded regions have the same meaning as in the Fig.~\ref{fig::pres}. The dash-dotted line shows the results of for the chiral condensate behavior with only pions included.}
	\label{fig::qq}	
\end{figure}

In Fig.~\ref{fig::qq} we show the resulting contributions to the chiral condensate. We separately show the contribution from the ``pole'' pressure and the LD contribution to the pressure to the total chiral condensate normalized by its vacuum value $\langle \bar q q \rangle = -(0.251 \gev)^3$.
Both of these contributions are negative, which reflects the fact that for a given temperature the pion mass increases with an increase of the current quark mass, therefore giving rise to the negative $\partial P_\pi / \partial m_0$. 
For comparison, we also show the pion contribution to the chiral condensate obtained in~\cite{Jankowski:2012ms} by the dash-dotted line. At low temperatures the $\qq_\pi^{\rm pole}$ gives the dominant contribution to the chiral condensate. We see that the low-temperature part of the pion contribution in our approach is lower, but is qualitatively the same as obtained in~\cite{Jankowski:2012ms}.

In the same temperature region $T \gsim 0.18 \gev \simeq 0.8\,T_c$ the LD region gives a noticeable contribution to the temperature dependence of the chiral condensate, comparable in magnitude with the $\qq_{\pi, \sigma}^{\rm pole}$. The magnitude of this contribution exhibits a maximum, corresponding to the maximum of the pressure in the Fig.~\ref{fig::pres}. Its temperature dependence at $T \lsim T_c$ is also qualitatively the same as that of the the $\qq$ in the MF approximation. At large temperatures it decays with the same rate as the pole contribution in line with what is described in subsect.~\ref{sect::thermo} for the pion pressure.
The total $\qq$ is then following the $\qq_{\pi, \sigma}^{\rm pole}$ at low temperatures. 

After calculating the perturbative contribution of the pions to the chiral condensate at finite temperature, we can define the ``new'' pseudocritical temperature $T_{c,\rm fl}^{\chi}$ of the chiral phase transition with an inclusion of both LD and QP mesonic contributions as
\begin{gather}
T_{c, \rm fl}^{\chi} = \arg \max \frac{\partial ( \langle\bar q q \rangle_{\rm MF} + \langle\bar q q \rangle_{\pi, \sigma}^{\rm QP} + \langle\bar q q \rangle_{\pi, \sigma}^{\rm LD})}{\partial T}.
\end{gather}
The inclusion of the pole contribution affects $T_{c, \rm fl}^{\chi}$ only a little, because the most rapid decrease at $T \to T_{c, \rm fl}^\chi$ comes from the quark-like contributions. In contrast with this, near $T_c^\chi$ the LD contribution contribution starts playing a noticeable role and enhances the chiral condensate decrease, as its contribution to the thermodynamics for $T \lsim T_c^\chi$ is qualitatively the same as that of the quark MF contribution. In the model with $T_0 = 0.27 \gev$ the inclusion of the LD term shifts the pseudocritical temperature to $T_{c, \rm fl}^{\chi}  = [227, 223]\mev$, corresponding to $\Lambda^{\rm LD} = [1, 2]\,\Lambda$ in comparison with $229\mev$ in the MF approximation to the PNJL model. This reduction is rather small but nevertheless important in the context of bringing the pseudocritical temperature of a constituent quark model closer to the observed in the lattice QCD.

\begin{figure}
	\includegraphics[width=.9\linewidth]{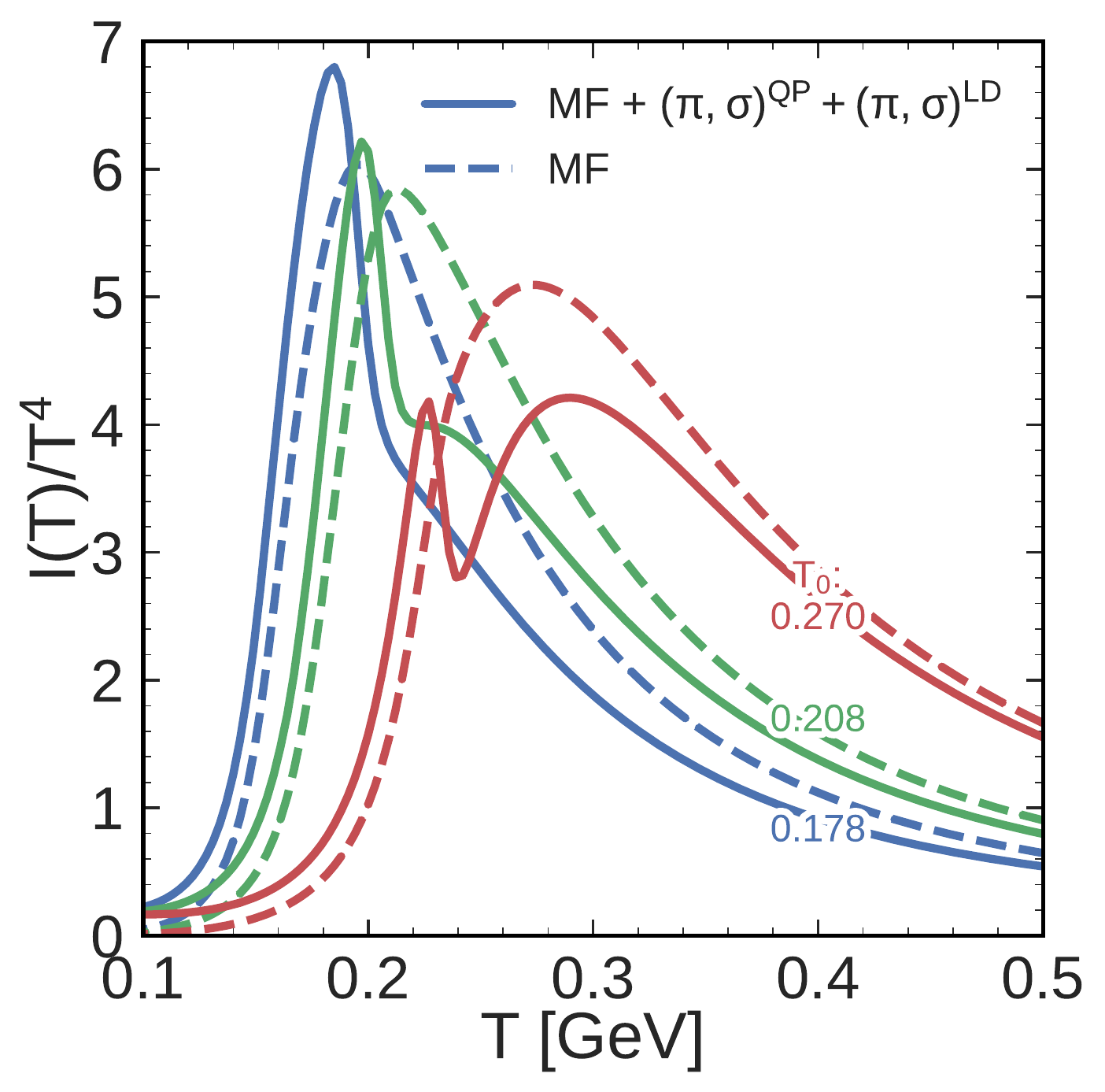}
	\caption{Scaled trace anomaly with the pion QP and LD contributions (solid lines, $\Lambda^{\rm LD} = \Lambda$) and the MF result (dashed lines) for $T_0 = [0.178, 0.208, 0.27]\gev$ labeled on the lines in \gev.}
	\label{fig::I_varT0}
\end{figure}

\subsection{Rescaling the Polyakov loop potential critical temperature parameter $T_0$}
\label{sect::varT0}
A rescaling of the Polyakov loop potential parameter $T_0$ to lower values is often used for comparing the PNJL results with the lattice QCD ones~\cite{Ratti:2005jh, Torres-Rincon:2017zbr}. Such a simple replacement is supported by the FRG calculations~\cite{Haas:2013qwp}, which have shown that the change of the Polyakov-loop potential due to the quark backreaction onto the Polaykov-loop potential can be approximated by with a similar rescaling of $T_0$. In this section we compare the results for the LD pressure between $T_0 = [0.27, 0.208, 0.178]\,\gev$, corresponding to pure YM and the values obtained in~\cite{Schaefer:2007pw} for 2 and 3 flavors, respectively. For brevity, we omit the results of the pressure and study only the trace anomaly as a more sensitive quantity for the pion LD contribution than the pressure.

In Fig.~\ref{fig::I_varT0} we show the trace anomaly for $T_0 = [0.27, 0.208, 0.178]\,\gev$. The dashed lines show the MF contribution to $I(T)$, and the solid lines are the total $I(T)$ with contributions from LD and QP regions of both $\pi$ and $\sigma$ excitations. For clarity we show only the $\Lambda^{\rm LD}=\Lambda$ case. We see that with a decrease of $T_0$ the overall magnitude of the LD contribution to $I(T)$ increases. As a consequence, the increase of the overall magnitue of the LD contribution with a decrease of $T_0$ makes the wiggle in the $\Lambda_{\rm LD} = \Lambda$ curve disappear, since the LD contribution contributes positively to the $I(T)$ maximum at lower temperature and negatively to the second one. For the cases $T_0 = [0.208, 0.178]\gev$ the presence of the pion contribution leads to a decrease of the location of the characteristic maximum of $I(T)$ from $[0.212, 0.197] \gev$ in the MF case to $[0.197, 0.185]\gev$ with the excitation contributions included.


The reason for this is the well-known increase in the separation of $T_c^\Phi$ and $T_c^\chi$ in the $T_0$ rescaling is used. This is illustrated in Fig~\ref{fig::MT}. For instance, in the case of $T_0 = 0.178\gev$ the $T_c^\Phi$ is $28\mev$ smaller than $T_c^\chi$. This leads to an increase of the quark distribution function at lower temperatures than in the case $T_0 = 0.27\gev$, as can be seen by the $\Phi(T)$ dependence in Fig.~\ref{fig::MT}. 
This entails a decrease of the quark effective mass and the subsequent increase of the LD contribution, as it corresponds to the emission (absorption) of pions by the quark medium.

For the case $T_0 = 0.178\gev$ we have evaluated the shifted pseudocritical temperatures $T_{c, \rm fl}^\chi$ as it was done in section~\ref{sect::qq}. The results are collected in Table~\ref{tab::Tcs}. In the case $T_0 = 0.178\gev$ we have obtained $T_{c, \rm fl}^\chi = [0.187, 0.183]\gev$ with $\Lambda^{\rm LD} = [1, 2]\,\Lambda$, respectively, which demonstrates a stronger reduction of the pseudocritical temperature than in the case $T_0 = 0.27\gev$. This demonstrates again the enhancement of the LD contribution in the case of a larger separation between the chiral and deconfinement transition preudocritical temperatures, and its importance if the goal is set to build a quantitative description of the QCD pseudocritical temperature at $\mu=0$ within a chiral quark model.

\begin{figure}
	\centering
	\includegraphics[width=.95\linewidth]{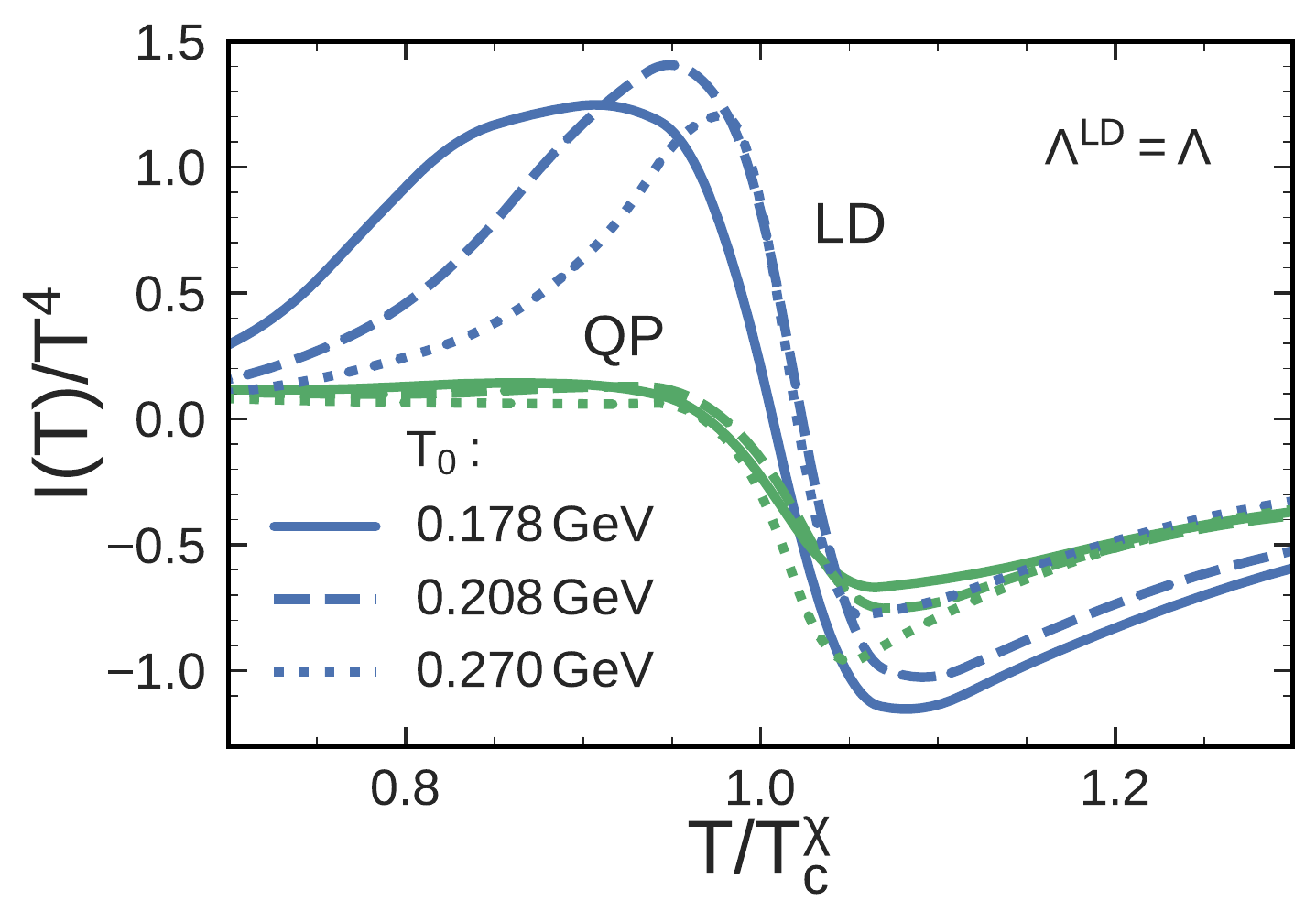}
	\caption{Quasipole and LD contributions with $\Lambda^{\rm LD} = \Lambda$ to the scaled trace anomaly for $T_0 = [0.178, 0.208, 0.27]\gev$ as functions of $T$ over the respective $T_c^\chi$, as collected in Table~\ref{tab::Tcs}. }
	\label{fig:I_varT0_pions}
\end{figure}


\section{Conclusion and outlook}
We have studied the 2-flavor PNJL model at finite temperature for a baryon chemical potential $\mu = 0$ in the mean-field approximation including the contributions to the pressure from the from dynamical quark current correlations in pion and sigma channels. The calculation of the meson pressure was performed using the generalized Beth-Uhlenbeck expression~\cite{Hufner:1994ma, Blaschke:2013zaa, Blaschke:2014zsa} through the quark-antiquark scattering phase shifts in the pion and sigma channels, which, as opposed to the standard Beth-Uhlenbeck approach, are medium-dependent and encode the Mott dissociation of the mesons. In this work, we paid particular attention to the effect of including the full momentum dependence of the pion phase shift as a function of frequency $\om$ and momentum $q$, including the Landau damping (LD) kinematic region with $s = \omega^2 - q^2 < 0$ on the thermodynamic quantities. 
In the LD region, the values of the spectral function and the phase shift are normally rather small and sometimes excluded out of consideration in similar models~\cite{Yamazaki:2013yua, Torres-Rincon:2017zbr}. In this paper, we have demonstrated that the inclusion of the LD region of the meson spectral strength to the calculation of the thermodynamic functions in fact leads to a significant contribution to the pressure centered around the chiral pseudocritical temperature $T_c^\chi$ because of the thermal enhancement of the low-frequency region by the Bose-Einstein distribution, and can affect the characteristic temperatures of the phase transition in the PNJL model.

Using the analytical expression of the 1-loop polarization operator in the pion channel, we have illustrated the thermal enhancement of the low-frequency mesonic excitations with finite momentum on the example of the pion structure factor. Such an increase leads to a significant contribution to the pressure integrand in the Beth-Uhlenbeck formula and a corresponding noticeable contribution to the pressure. Below $T_c^\chi$, the pressure originating from the LD region grows with the same rate as the quark contribution, since LD corresponds to the emission and absorption of pions by the quark thermal bath. The analytic expressions for the LD polarization loop for non-relativistic quarks contain a similar temperature exponent $\exp({-\sqrt{(2 m)^2 + q^2} / 2 \, T})$, with the same dependence on the quark effective mass $m$ as the quark pressure, but with the large-momentum tail regulated by twice the medium temperature. This is partially responsible for the significant magnitude of the LD contribution. We have varied the threshold in the LD region $\Lambda^{\rm LD} = [1-2]\,\Lambda$, where $\Lambda$ is the quark 3-momentum cutoff which regularized the PNJL vacuum pressure, and have found a sizeable 4-fold enhancement of the result for the LD pressure for $T \gsim T_c^\chi$. For the case of the NJL model we could obtain a semi-analytic estimate in the non-relativistic region, given in the Appendix~\ref{app::analytic}, which confirms the numerical results in the corresponding applicability range of temperatures.
This contribution might be responsible for the kinks in the pressure in~\cite{Roessner_arxiv}, which are not seen in the published version~\cite{Roessner:2007gha} where only the quasipole contribution was left in the mesonic pressure.The LD contribution to the pressure leads to a change in the trace anomaly which is more sensitive to any additional contribution since at $\mu = 0$ it is proportional to the temperature derivative of the pressure. The LD contribution to the trace anomaly changes sign at $T \simeq T_c^\chi$, because the pion contribution for $T \gsim T_c^\chi$ is decreased due to the pion dissociation.

We have studied the contribution of the meson gas to the temperature dependence of the chiral condensate melting, calculated perturbatively without iterating the self-consistency condition, by means of the Hellmann-Feynman theorem. The resulting chiral condensate at low temperatures exhibits a decrease due to the pion pole excitations only. As the temperature approaches $T_c^\chi$, the contribution of the LD region becomes important, which qualitatively for $T \lsim T_c^\chi$ has the same temperature dependence as the quark contribution. The inclusion of such a term allows to make $T_c^\chi$ lower by several MeV depending on the choice of $\Lambda^{\rm LD}$.

Finally, we have studied the effect of lowering the $T_0$ parameter of the Polyakov-loop potential, suggested by the renormalization-group arguments~\cite{Schaefer:2007pw,Haas:2013qwp}, on the LD contribution to the thermodynamic quantities. By the example of the trace anomaly, as the most sensitive quantity, we have shown that decreasing $T_0$ leads to an overall increase of the LD contribution to the thermodynamics. This is a consequence of the well-known fact of separation of $T_c^\chi$ and $T_c^\Phi$ under a decrease of $T_0$, which leads to a relative increase of the quark distribution function for $T \lsim T_c^\chi$. In turn, this enhances the LD contribution, because it corresponds to emission/absorption of pions by quark thermal bath and is therefore sensitive to the quark distribution function.
All these effects described above are expected to be present not only in the mean-field-based approaches to the PNJL model but also in those including the $1/N_c$ corrections to the quark propagator if the mesonic polarization operators are evaluated with quark quasiparticle propagators at the mean-field level, as was done in this work.

This work is limited to $N_f = 2$ case and focuses only on the pion contribution, but after generalization to $N_f = 2 + 1$ even more meson degrees of freedom would give rise to a similar contribution to the thermodynamics. 
Moreover, a self-consistent solution of the gap equation and the equation of motion for the Polyakov loop variable is necessary and will be reported elsewhere. Physically the Landau damping corresponds to a non-dissipative process of the energy transfer from meson system to the quark subsystem and its inverse, and therefore the inclusion of the quark back-reaction is necessary to provide reliable conclusions. However, the soft quark excitations are not enhanced by the thermal Bose factor, and we expect our conclusions about the mesonic contribution to be qualitatively unchanged. The inclusion of the back-reaction can be done within a self-consistent, e.g. $\Phi$-derivable scheme, which is a subject of our future work.



%
%
%
%
%
%
\subsection*{Acknowlegdements}
We thank Krzysztof Redlich, Chihiro Sasaki, Pok Man Lo, Evgeny Kolomeitsev and Dmitry Voskresensky for the discussions and their interest in this work. The work benefited a lot from numerous conversations and help from Oleksii Ivanytskyi. 
K.A.M. acknowledges the hospitality of the University of Wroclaw, where the most part of this work was done. The work of K.A.M. was supported by the Foundation for the Advancement of Theoretical Physics and Mathematics “BASIS”, project {№17-15-568-1}.
This work has been supported in part by the Polish National Science Centre (NCN) under grant No. 2019/33/B/ST9/03059.
\vspace{10mm}

\appendix
\section{Expressions for the meson polarization operator}
\label{app::meson_exprs}

The imaginary part of the polarization operator of the meson $M = (\pi, \sigma)$ can be expressed for PNJL model in the following compact form~\cite{Hatsuda:1994pi,Roessner_arxiv,Blaschke:2013zaa}:
\begin{widetext}
	\begin{gather}
		\Im \Pi_M (\om, q) = -\kappa_M \frac{N_c N_f}{8 \pi q} \Big(\theta(s - 4 m^2)\Big[\theta(\om) J^+_{\rm pair}(\om, q) + \theta(-\om) J^-_{\rm pair}(\om, q) \Big] + \theta(-s) J_{\rm LD}(\om, q)\Big), \,\, \\ 		J_{\rm pair}^\pm(\om, q) = T \ln\Big( \frac{\phi^\mp(E_-) \phi^\mp(-E_-)}{\phi^\mp(E_+) \phi^\mp(-E_+)}\Big), \,\, J_{\rm LD}(\om, q) = T \ln \Big(\frac{\phi^+(E_-) \phi^-(E_-)}{\phi^+ (- E_+) \phi^-(-E_+)}\Big), 
	\end{gather}
\end{widetext}
where $m$ in the constituent quark mass, $s = \om^2 - q^2, \,\kappa_\pi = s, \,\kappa_\sigma = s - 4 m^2$, $E_\pm = \frac{\om}{2} \pm \frac{q}{2}\sqrt{1 - \frac{4 m^2}{\om^2 - q^2}}$, and the distribution function is defined as
\begin{align}
	\phi^\pm(\eps) = 
		\Big(\dfrac{1}{y_\pm^3 + 3y_\pm(\bar \Phi + \Phi y_\pm) + 1}\Big)^{1/3},
\end{align}
where $y_\pm = e^{(\eps \pm \mu)/T}$. The NJL expressions can be restored by taking the limit $\Phi, \bar\Phi \to 1$. The real part is then calculated as 
\begin{gather}
	\Re\Pi_M(\om, q; T) = - \int\limits_0^{\Lambda_4^2(q)} \frac{d s'}{\pi} \frac{\Im \Pi_M(\sqrt{s'},  q; T)}{\omega^2 - s'}, 
\end{gather}
where $q = |\vec q|$, and it is necessary to use the mass- and momentum-dependent 4-dimensional cutoff $\Lambda_4^2( q) = 4 (\Lambda^2 + m^{2}) +  q^{\,2}$ in order to keep thermodynamic consistency~\cite{Hatsuda:1994pi}.

For $\mu = 0$ the LD part in the NJL model has the following asymptotic expressions:
\begin{gather}
 \quad J_{\rm LD}(\om, q; T) \underset{T \to 0}{\longrightarrow} 
	-4 T \exp(-\frac{q}{2 T} \sqrt{1 + \frac{4m^2}{q^2 - \om^2}}) \nn \times \sinh{\frac{\om}{2 T}}
	\underset{\om \to 0}{\longrightarrow} -2 \om \exp({-\dfrac{\sqrt{{4 m^2+q^2}}}{2 T}}).
	\label{eq::J_LD_0}
\end{gather}
The dependence of the LD part on the constituent quark mass over the temperature contains therefore the same factor $e^{-m/T}$ as the quark contribution, while the large-momentum tail has an effective temperature of $2 T$. The factor $\omega$ in the low-frequency limit eliminates the pole of the Bose-Einstein distribution in the integrals in Eq. \eqref{eq::fp}. A more general expression for the polarization operator in the LD limit, which includes the PNJL case and is valid for any $T$ as long as $\omega \ll q$, is~\cite{Yamazaki:2012ux}
\begin{gather}
	\Im\Pi_\pi^{\rm LD}(\om, q; T) \simeq N_c N_f \frac{\om^2 - q^2}{4 \pi} \frac{\om}{q} f_\Phi(\frac{q}{2} \xi), 
	\label{eq::J_LD_ollq}
\end{gather}
with $f=f_\Phi^+ = f_\Phi^-$ for $\mu = 0$ is given by Eq.~\eqref{eq::f_phi}. This expression also proves to be rather accurate for nearly all $0 < \om \lsim q$, if $q \lsim 2 T$. In the limit $\Phi \to 1, \bar \Phi \to 1$ the NJL case is restored, as $f_\Phi$ becomes a usual Fermi-Dirac distribution.

\section{Low-temperature analytic estimate of the LD contribution to the pressure}
\label{app::analytic}
The asymptotic expression~\eqref{eq::J_LD_ollq} for $\Im\Pi_\pi(\om, q)$ allows to estimate the parametric dependence of the LD pressure analytically.
Concerning the $\Re\Pi_\pi(\om, q)$, in the non-relativistic case, corresponding to the best applicability range of~\eqref{eq::J_LD_ollq}, we can use the pole approximation for the real part of the inverse pion propagator
\begin{gather}
	\Re(\frac{1}{2 G_s} + \Pi_\pi) \simeq -\frac{1}{g_{\pi q q}^2} \frac{1}{\omega^2 - m_\pi^2(T) - q^2}, \nn
	g_{\pi q q}^2 = -\Big(\frac{\partial \Re\Pi_\pi(\om, q=0, T)}{\partial \om^2} \Big|_{\omega = m_\pi} \Big)^{-1},
	\label{eq::rePi_approx}
\end{gather}
where $m_\pi(T)$ is the pion mass, determined by Eq.~\eqref{eq::m_pi}, and $g_{\pi q q}^2 (T)$ is the squared pion-quark coupling~\cite{Klevansky:1992qe, Hatsuda:1994pi}.
The phase shift~\eqref{eq::delta_pi} in the LD region does not get close to $\pi$ and we can approximate it by the argument of the $\arctan$. Then the pion pressure reads
\begin{gather}
	P_\pi \simeq \frac{N_c N_f d_\pi}{8 \pi^4} \int\limits_{0}^\Lambda q dq \int\limits_0^q \om d\om \frac{|s|}{m_\pi^2 + |s|}  \dfrac{1}{e^{\frac{q}{2T} \xi(\om, q)} + 1} \nn \times \frac{1}{\exp(\om/T) - 1}, \,\, s = \omega^2 - q^2, \,\, \xi = \sqrt{1 + \frac{4 m^2}{|s|}},
\end{gather}

In order to factor out the essential temperature dependence, we replace the sharp momentum cut-off by the dipole-type form-factor  $L(\alpha, x) = \alpha^2/(\alpha^2 + x^2)$ and change variables to $q = T x \cosh \chi, \, \om = T x \sinh \chi \equiv |s| \cdot y$ to transform the integration limits to the infinite quarter-plane. After further non-relativistic expansions in terms of new variables, we arrive to an (underestimating) expression
\begin{gather}
	P^{\rm LD}_{\pi, \rm NR} = \frac{N_c N_f d_\pi g_{\pi q q}^2}{8 \pi^4} \frac{m^2}{m_\pi^2} T^4 e^{-m/T} \times \nn F(\frac{\Lambda_\pi}{\sqrt{mT}}, \frac{m_\pi}{\sqrt{mT}}, \frac{m}{T}),
\end{gather}
where $d_\pi = 3$ is the pion isospin degeneracy factor, and the dimensionless function $F$ is defined as
\begin{gather}
	F(\alpha, \beta, \gamma) = \int\limits_{0}^{\infty} dx x^3 e^{-x^2/8} L(\alpha,x) L(\beta, x) G(\frac{1}{x^2} + \frac{1}{8 \gamma}), \nn G(\beta) = \int\limits_0^\infty dy \frac{y e^{-\beta y^2/2}} {e^y - 1}. \nonumber
\end{gather}

\bibliography{biblio_clean.bib}

\end{document}